\documentclass[superscriptaddress, twocolumn, prl]{revtex4-1}
\usepackage[T1]{fontenc}
\usepackage{hyperref}
\usepackage{amsmath}
\usepackage{amsthm}
\usepackage{amssymb}
\usepackage{graphicx}
\usepackage{color}
\usepackage{qcircuit}
\usepackage{braket}
\usepackage{comment}
\usepackage{bbm}
\usepackage{bm}

\hypersetup{colorlinks,linkcolor={blue},citecolor={magenta},urlcolor={blue}}

\newcommand{\beq}{\begin{equation}}
\newcommand{\eeq}{\end{equation}}
\definecolor{Pr}{rgb}{0.4,0.3,0.9}

\newcommand{\spliteq}[1]{\begin{equation}
\begin{split}
#1
\end{split}
\end{equation}
}

\def\ket#1{\mathop{|#1\rangle}}
\def\bra#1{\mathop{\langle #1|}}

\definecolor{JM}{RGB}{4,116,149}

\begin{document}
%\title{Quantum Approximate Continuous Optimization}
%Alternate title
\title{A Quantum Approximate Optimization Algorithm for continuous problems}
\author{Guillaume Verdon}
\email{gverdona@uwaterloo.ca}
\affiliation{Department of Applied Mathematics, University of Waterloo, Waterloo, Ontario, N2L 3G1, Canada}
\affiliation{Institute for Quantum Computing, University of Waterloo, Waterloo, Ontario, N2L 3G1, Canada}
\affiliation{Perimeter Institute for Theoretical Physics, Waterloo, Ontario, N2L 2Y5, Canada}

\author{Juan Miguel Arrazola}
\email{juanmiguel@xanadu.ai}
\affiliation{Xanadu, 777 Bay Street, Toronto, Ontario, M5G 1S5, Canada}

\author{Kamil Br\'adler}
\affiliation{Xanadu, 777 Bay Street, Toronto, Ontario, M5G 1S5, Canada}

\author{Nathan Killoran}
\affiliation{Xanadu, 777 Bay Street, Toronto, Ontario, M5G 1S5, Canada}

\begin{abstract}
We introduce a quantum approximate optimization algorithm (QAOA) for continuous optimization. %This approach can be used for black-box optimization of both continuous and discrete problems. 
The algorithm is based on the dynamics of a quantum system moving in an energy potential which encodes the objective function. By approximating the dynamics at finite time steps, the algorithm can be expressed as alternating evolution under two non-commuting Hamiltonians. We show that each step of the algorithm updates the wavefunction in the direction of its local gradient, with an additional momentum-dependent displacement. For initial states in a superposition over many points, this method can therefore be interpreted as a coherent version of gradient descent, i.e., `gradient descent in superposition.' This approach can be used for both constrained and unconstrained optimization. In terms of computational complexity, we show how variants of the algorithm can recover continuous-variable Grover search, and how a single iteration can replicate continuous-variable instantaneous quantum polynomial circuits. We also discuss how the algorithm can be adapted to solve discrete optimization problems. Finally, we test the algorithm through numerical simulation in optimizing the Styblinski-Tang function.
\end{abstract}
\maketitle
\textit{Introduction---} Variational quantum circuits have emerged as a new paradigm in quantum algorithmic development. In this setting, quantum algorithms are built by designing specific circuit architectures and subsequently choosing their gate parameters via classical optimization techniques~\cite{mcclean2016theory}. Circuit architectures can be chosen to meet the constraints of existing hardware, while the classical parameter optimization serves the purpose of delegating part of the computational burden and implicitly adapting to errors present in the circuit~\cite{kandala2017hardware, wecker2015progress, yang2017optimizing}. Variational quantum circuits have been proposed for chemistry calculations \cite{peruzzo2014variational}, factoring~\cite{anschuetz2018variational}, implementing Grover's search algorithm~\cite{morales2018variationally}, quantum autoencoders~\cite{romero2017quantum}, state diagonalization~\cite{larose2018variational}, and quantum neural networks \cite{verdon2017quantum,farhi2018classification, chen2018universal, killoran2018continuous, verdon2018universal}.

Of particular interest are applications to optimization. The term quantum approximate optimization algorithm (QAOA) is often used to refer to this class of variational quantum algorithms, in reference to  seminal work by Farhi et al.~\cite{farhi2014quantum}. There are active efforts to study the performance of QAOA and to develop improvements and extensions~\cite{farhi2016quantum,farhi2017quantum,moll2017quantum, hadfield2017quantum, crooks2018performance, bapat2018bang, zhou2018quantum, crooks2018performance, fingerhuth2018quantum, brandao2018fixed}. QAOA has currently only been considered in the context of discrete optimization, yet many problems arising in finance~\cite{cornuejols2006optimization}, machine learning~\cite{sra2012optimization},  and engineering~\cite{rao2009engineering} fall under the class of continuous optimization, where the goal is to minimize a real-valued function over a continuous domain. To enable quantum optimization algorithms to have a wider impact in these areas, it is necessary to extend the applicability of existing methods.

We introduce a quantum approximate algorithm for continuous optimization. The algorithm follows the alternating operator approach that characterizes QAOA: the objective function is encoded in a cost Hamiltonian and the algorithm proceeds by interchanging steps of time evolution under two non-commuting operators: the cost Hamiltonian and the mixer Hamiltonian. The time intervals of each step are the variational parameters to be classically optimized. We show that for appropriate choices of the mixer Hamiltonian, a single step of the algorithm mimics the gradient descent methods developed in deep learning: the starting point is updated in the direction of the gradient of the cost function, with the addition of a momentum-dependent displacement. By setting an initial state in superposition over several starting points, the algorithm can be interpreted as performing gradient descent in superposition. Indeed, the resulting evolution is equivalent to the quantum dynamics of a particle moving in the presence of the potential describing the cost function. We highlight the versatility of the algorithm by discussing how discrete and constrained optimization problems can be tackled in this setting. Additionally, we show that the algorithm allows Grover-like quadratic speedups for search problems and, under standard complexity assumptions, can sample from distributions that cannot be reproduced in classical polynomial time.

An ongoing challenge in the implementation of QAOA algorithms is to find efficient methods of optimizing the variational parameters of the circuits \cite{brandao2018fixed,zhou2018quantum}. In our case, it is possible to choose the gate parameters to reproduce well-known heuristics of gradient descent algorithms used in deep learning, which avoid overshooting and permit the algorithm to settle in local minima. This feature can significantly reduce the difficulty of optimizing classical parameters. We illustrate the behaviour and properties of the algorithm through numerical experiments, finding that for simple optimization tasks, parameters can be set according to gradient descent heuristics without the need for sophisticated optimization algorithms. 

\textit{Quantum algorithm--- }
For simplicity, we describe the algorithm in the model of continuous-variable quantum computing, where registers are quantum harmonic oscillators characterized by position $\hat{x}$ and momentum $\hat{p}$ operators \cite{lloyd1999quantum, braunstein2005quantum}. However, the algorithm can also be implemented in the standard qubit model, as was studied in Ref. \cite{verdon2018universal}, where registers are digitally simulated quantum harmonic oscillators.

Consider a function of $N$ variables $f(\bm{x}): \mathbb{R}^N \mapsto \mathbb{R}$. A continuous optimization problem consists of finding an approximate minimum $\bm{x}^\ast$ such that $f(\bm{x}^\ast)\approx \min_{\bm{x}\in \mathbb{R}^n} f(\bm{x})$. This can be equivalently phrased as finding a state that approximately minimizes the expectation value $\braket{\hat{H}_C}$ of a \emph{cost Hamiltonian} $\hat{H}_C= f(\bm{\hat{x}})$. Following the alternating operator ansatz \cite{hadfield2017quantum}, we define also a \emph{mixer Hamiltonian} that does not commute with $\hat{H}_C$. We first focus on what we call the \textit{kinetic mixer} $\hat{H}_M=\tfrac{1}{2}\sum_{j=1}^N\hat{p}_j^2:=\tfrac{1}{2}\bm{\hat{p}}^2$, but other choices are also possible. In the Heisenberg picture, evolving under the mixer Hamiltonian, the position operator is transformed as
\begin{equation}\label{eq:parshift}
    e^{i\gamma\bm{\hat{p}}^2/2}   \, \bm{\hat{x}} \,
    e^{-i \gamma\bm{\hat{p}}^2/2} =\bm{\hat{x}} + \gamma \bm{\hat{p}}.
\end{equation}
Conjugately, evolving under the cost Hamiltonian generates a translation in momentum of the form
\begin{equation}\label{eq:momshift}
    e^{i\eta f(\bm{\hat{x}})}    \bm{\hat{p}}e^{-i\eta f(\bm{\hat{x}})} = \bm{\hat{p}}- \eta\,\bm{\nabla}\! f(\bm{\hat{x}}) ,
\end{equation}
i.e., the momentum is shifted by the negative gradient of the cost function. This formula can be easily verified to be valid for any analytic function $f$. Alternating evolution under the cost and mixer Hamiltonians leads to the transformation
\spliteq{\label{eq:double_upt}
       \bm{\hat{x}}&\rightarrow \bm{\hat{x}} +\gamma\bm{\hat{p}}- \eta\gamma\, \bm{\nabla}\! f(\bm{\hat{x}}).
}
This update rule has a direct analogy with gradient descent with momentum: each part of the wavefunction is updated by descending in the direction of its local gradient, with an additional momentum-dependent displacement. For infinitesimal times $\gamma,\eta\sim \mathrm{d}t$, this change is equivalent to the quantum dynamics of a particle undergoing motion in a high-dimensional potential $f(\bm{x})$.

The algorithm is defined as follows. Given an initial state $\ket{\Psi_0}$, we apply the QAOA unitary
\begin{equation}\label{eq:QAOA_ans}
    \hat{U}(\bm{\eta},\bm{\gamma}) =\prod_{j=1}^P e^{-i\bm{\gamma}_j \hat{H}_M} e^{-i\eta_j \hat{H}_C},
\end{equation}
to produce the output state $\ket{\Psi_{\bm{\eta},\bm{\gamma}}}= \hat{U}(\bm{\eta},\bm{\gamma})\ket{\Psi_0}$, where \(\bm{\eta} = \{\eta_j\}_{j=1}^P\) and \(\bm{\gamma} = \{\gamma_j\}_{j=1}^P\).
Each register is then measured in the position basis to reveal a sample point $\bm{x}$ with probability proportional to  $|\!\braket{\bm{x}|\Psi_{\bm{\eta},\bm{\gamma}}}\!|^2$, where $\{\ket{\bm{x}}\}_{\bm{x}\in\mathbb{R}^n}$ are the eigenstates of $\bm{\hat{x}}$. This process can then be repeated, enabling an iterative search for additional approximate solutions. 

Samples can also be used as input to classical optimization algorithms to decide how to update the variational parameters $\bm{\eta,\gamma}$ \cite{farhi2014quantum, verdon2017quantum, otterbach2017unsupervised, zhou2018quantum, crooks2018performance}. In our case, however, the role of the parameters is transparent: $\bm{\gamma}$ determines the strengths of the shifts in momentum, while $\bm{\eta}$ plays the role of the learning rates at each iteration, deciding the amount of displacement along the local gradient. It is therefore possible to choose the parameters in accordance with established heuristics of gradient descent algorithms -- for instance, a gradual reduction of both the learning rate and momentum boosts with each step of descent -- allowing the algorithm to settle in local minima. Initial parameters can be chosen using standard principles from gradient descent, then fine-tuned through classical parameter optimization.

Finally, it is in principle possible to choose different mixer Hamiltonians as long as they do not commute with the cost Hamiltonian. A concrete example is the number mixer $\hat{H}_M=\bm{\hat{n}} = \sum_{j=1}^N \hat{a}_j^\dagger \hat{a}_j$, where $\hat{a}_j$ is the annihilation operator of mode $j$. In this case, the update rule after a single step satisfies
\(
\bm{\hat{x}}\rightarrow \bm{\hat{x}} +\gamma\bm{\hat{p}}- \eta\gamma \bm{\nabla} f(\bm{\hat{x}})+\mathcal{O}(\gamma^2),
\)
reproducing the result for the kinetic mixer up to small corrections in $\bm{\gamma}$. One advantage of this mixer is that it is readily implementable using phase shifters \cite{weedbrook2012gaussian}. As for the kinetic mixer, it can be applied as a product of independent Gaussian transformations on each mode \cite{weedbrook2012gaussian, su2018implementing}. In order to compile the exponentials of the cost Hamiltonian, in the cases where the function $f(\bm{x})$ is a known polynomial, the unitary $e^{-i\eta f(\bm{\hat{x}})}$ can be decomposed following the gate decomposition methods of \cite{sefi2011decompose, kalajdzievski2018exact}. If the function is a black box, then it must be queried in superposition in order to create a phase oracle \cite{gilyen2019optimizing}. In principle, one possibility is to use a quantum neural network to learn a functional approximation of $f(\bm{x})$ in an interval $\bm{x}\in[a,b]^n$ \cite{killoran2018continuous,arrazola2018machine} and use quantum phase kick backpropagation \cite{verdon2018universal} to enact $e^{-i\eta f(\bm{\hat{x}})}$  on the space $\mathrm{span}\{\ket{\bm{x}}\, :\, \bm{x}\in[a,b]^n\}$. Overall, the resulting circuit depth will depend on the specific properties of $f(\bm{x})$. The smallest circuit depths will result when $f(\bm{x})$ can be well approximated by a low-degree polynomial in the region of interest. In principle, this region of interest can be adaptively shifted during the quantum-classical optimization iterations of the algorithm such that most of the probability measure of the output wavefunction, as determined from multiple measurements, remains within the said domain. The input state's position and squeezing levels can be adjusted accordingly to fit within each new domain of polynomial approximation. 

\textit{Constrained Optimization---}
In addition to unconstrained continuous optimization, the algorithm can be applied to continuous constrained optimization problems, which are of relevance in many areas \cite{neumaier2004complete,pinter2013global,storn1997differential}.
A general constrained optimization problem consists of finding the minimum of a function $\min_{\bm{x}\in\mathbb{R}^n}f(\bm{x})$ subject to a set of equality constraints, \(\{g_i(\bm{x}) = c_i\}_{i=1}^n \), and a set of inequality constraints, \(\{h_j(\bm{x}) \geq d_j\}_{j=1}^m\). To enforce these constraints during the optimization, we can add energetic penalties to the cost Hamiltonian, $\hat{H}_C \mapsto f(\bm{\hat{x}}) + V_E(\bm{\hat{x}}) + V_I(\bm{\hat{x}})$. Here, $V_E$ and $V_I$ are the constraining potentials for the equality and inequality clauses. As Eq.~\eqref{eq:double_upt} shows, the gradient of the potential induces shifts of the position of the wavefunction, thus the key is to add potentials whose gradients drives the wavefunction towards the region obeying the various constraints. 

For the equality constraints, a possible choice of constraining potential is
\(
    V_E(\bm{\hat{x}}) = \sum_{i=1}^n\lambda\left( g_i(\bm{\hat{x}})-c_i\right)^2   
\)
with large $\lambda$. This potential creates a valley of low-energy values in the landscape which drives the dynamics towards the submanifold satisfying the equality constraints \(\{g_i(\bm{x}) = c_i\}\); this happens to be where the gradient vanishes $\nabla V_E(\bm{x}) = 0$. 
As for the inequality constraints, we take an analogous approach and use the constraining potential, \(
V_I(\bm{\hat{x}})=  \sum_{j=1}^m \mathcal{R}\left[\beta(d_j- h_j(\bm{\hat{x}}))\right]
\) where $\mathcal{R}(x)$ is an analytic approximation to the \textit{rectifier} function $R(x) \equiv \max{\{0,x\}}$, and $\beta$ is large. This rectifier function has a linear increasing slope for positive values and is flat for negative values, thus the $V_I$ potential will drive the dynamics of the optimization towards the region obeying the inequality constraints \(\{h_j(\bm{x}) \geq d_j\}_{j=1}^m\). One option for an analytic approximation is the \emph{swish} function \cite{ramachandran2017swish,ramachandran2018searching}: choosing $\mathcal{R}(x) =\xi(x) \equiv  x \cdot \sigma(x)$, where $\sigma(x)\equiv 1/(1+e^{-x})$ is the sigmoid function.
% Note that in the limit of large $\beta$, we recover a function proportional to the rectifier $\lim_{\beta\rightarrow \infty} \xi(\beta x)/\beta = R(x)$, in which case we effectively have a wall-like (steep slope) potential. Alternatively, one could choose $\mathcal{R}(x) =\zeta(x) \equiv \ln(1+e^{x})$, known as the \emph{softplus} function \cite{ramachandran2018searching}. Note that the gradient of the softplus is the sigmoid, $\zeta'(x)=\sigma(x)$, and that it also converges to the steep rectifier function in the limit of large $\beta$, $\lim_{\beta\rightarrow \infty} \zeta(\beta x)/\beta = R(x)$, thus yielding the desired constraining dynamics.

%Both alternatives could play the role of being a smooth ReLU-like potential, in the limit of large $\beta$ these become highly sloped and effectively form a steep wall in the potential landscape. The quadratic potential is more like a well which will concentrate the wavefunction onto the joint equality constraints submanifold.

\textit{Encoding discrete optimization problems---} The algorithm we have introduced is ideally suited for continuous optimization problems. However, it is versatile enough to handle even discrete optimization tasks. A broad class of discrete problems of interest in various fields is polynomial unconstrained binary optimization (PUBO) problems \cite{babbush2014construction,rieffel2015case,perdomo2017readiness}, which can be phrased as the task of finding the ground state of a Hamiltonian $\hat{H} = f(\bm{\hat{Z}})$ where $f$ is a polynomial function of the Pauli operators on $n$ qubits $\bm{\hat{Z}} = \{\hat{Z}_j\}_{j=1}^n$, generally of the form $
    f(\bm{\hat{Z}}) = \sum_{\bm{b}\in\mathbb{Z}_2^n} \alpha_{\bm{b}}\bm{\hat{Z}}^{\bm{b}}$ where $\bm{\hat{Z}}^{\bm{b}}:=\textstyle\bigotimes_{j=1}^n\hat{Z}^{b_{j}}_j$, $\alpha_{\bm{b}}\in \mathbb{R}$. The subclass of quadratic PUBO problems (so-called QUBO problems \cite{smelyanskiy2012near,venturelli2015quantum}) is of particular interest in a wide array of applications \cite{perdomo2017readiness}, including deep learning \cite{amin2018quantum,verdon2017quantum}. %Let us now outline our approach to encode such a discrete PUBO Hamiltonian as a potential over an $n$-mode continuous-variable system.

To encode the binary polynomial $f(\bm{\hat{Z}})$ as a function over the continuum, we consider the Pauli $\hat{Z}$ of each qubit observable to be effectively equivalent to the sign of the position observable of a corresponding quantum oscillator $\hat{Z}_j \leftrightarrow \text{sgn}(\hat{x}_j)$. %(notice these operators share the same spectrum).
As an analytic approximation to this sign function, we use the hyperbolic tangent, $\hat{Z}_j\mapsto \tau_\beta(\hat{x}_j) \equiv \tanh(\beta \hat{x}_j)$, where $\beta$ is a tunable parameter allowing for adjusting the sharpness of the step. % (notice we recover the sign function in the limit of large $\beta$).
The binary polynomial gets converted to $f(\bm{\hat{Z}}) \mapsto f(\bm{\tau}_\beta(\bm{\hat{x}}))$ where $\bm{\tau}_\beta(\bm{\hat{x}}) = \{\tanh(\beta \hat{x}_j)\}_j$. Although this prescription allows for encoding the cost function as plateaus of a potential in the continuum, we further add a constraining potential to keep the wavefunctions contained in a discrete lattice of wells. Specifically, we add to each oscillator a double-well potential of the form
$W_{\omega,\lambda}(\bm{\hat{x}})=\sum_j{\omega^2\over2}(\hat{x}_j^2-\lambda^2)^2$. Each oscillator then has a left ($x<0$) and right ($x>0$) well. We assign the states $\ket{1}$ and $\ket{0}$ to these wells respectively. %, as these correspond to eigenvalues $-1$ and $1$ of the the sign observable.
If one considers this potential on its own, the degenerate low-energy subspace of the potential forms an effective discrete lattice subspace; that of the wells of the potential. These wells prevent the wavefunctions from straying too far from the origin while keeping them in regions where the tanh well-approximates the sign function. The parameters $\omega$ and $\lambda$ can be adjusted to tune the separation of the wells and minimize cross-talk to a negligible level.

Putting all of these ingredients together, the prescription to encode a PUBO problem with binary polynomial $f(\bm{x})$ is to pick a cost Hamiltonian
\begin{equation}\label{eq:convert}
\hat{H}_C \equiv f\left(\bm{\tau}_\beta(\bm{\hat{x}})\right) +  W_{\omega,\lambda}(\bm{\hat{x}}),
\end{equation}
where the parameters $\omega,\lambda,\beta$ are to be tuned as described above. The mixer Hamiltonian $\hat{H}_M$ can be chosen to be the kinetic or number mixer, which we introduced previously.

\textit{Complexity--- } Quantum approximate optimization algorithms are, by construction, heuristics, and therefore their merits are ultimately decided by testing their performance on concrete problems. Nevertheless, here we describe two notable complexity-theory statements about the algorithm. First, we show that QAOA circuits can encode Grover's search algorithm over continuous spaces~\cite{pati2000quantum}, which is known to achieve a quadratic speedup compared to classical algorithms. Additionally, we outline how a single step of the algorithm can replicate continuous-variable instantaneous quantum polynomial (CV-IQP) circuits, which are believed to be impossible to efficiently simulate classically~\cite{douce2017iqp, arrazola2017quantum}. 
    
\begin{center}
\begin{figure*}[t!]
\begin{tabular}{cccccc}
\includegraphics[width=0.4 \columnwidth]{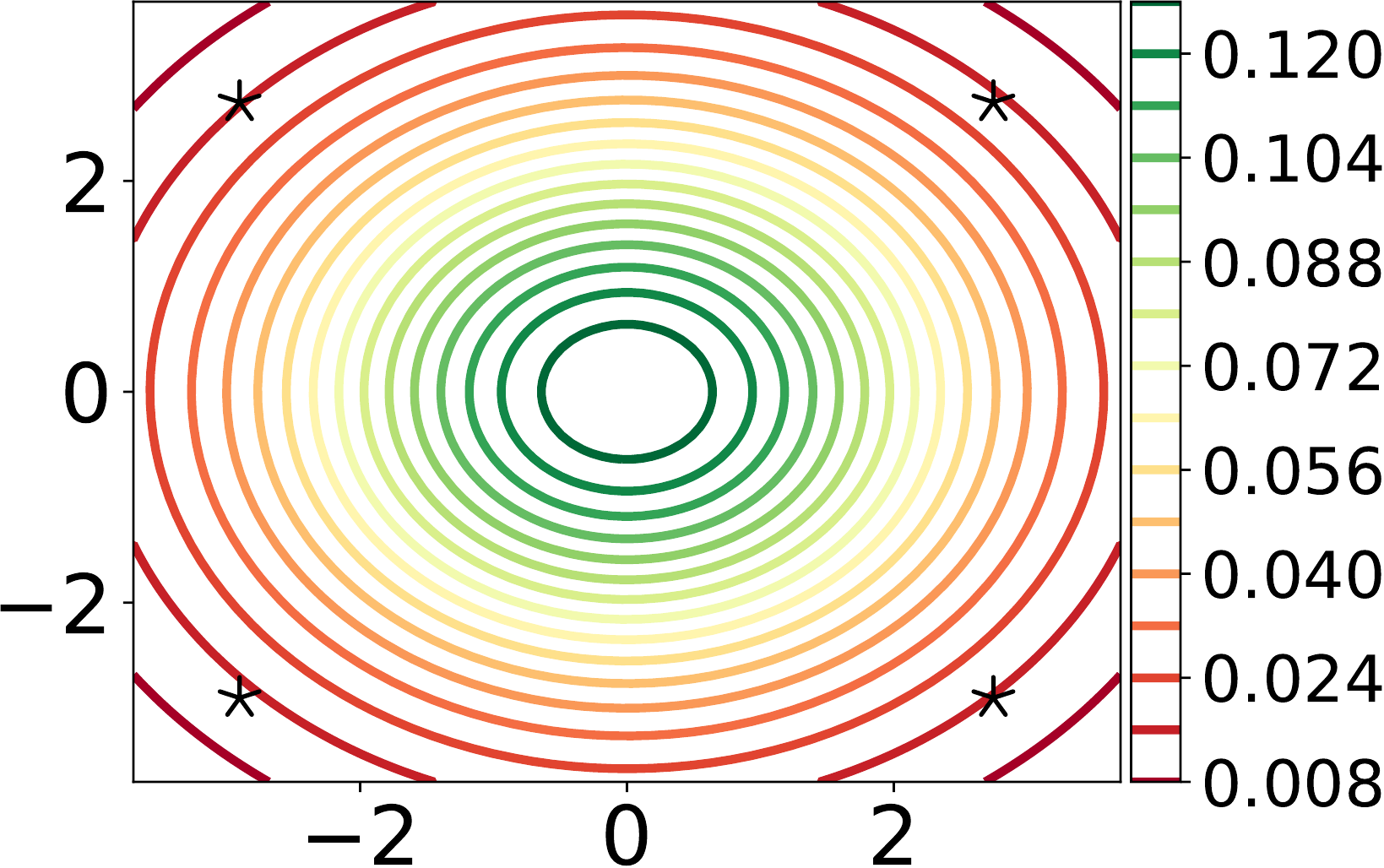}& \includegraphics[width=0.4 \columnwidth]{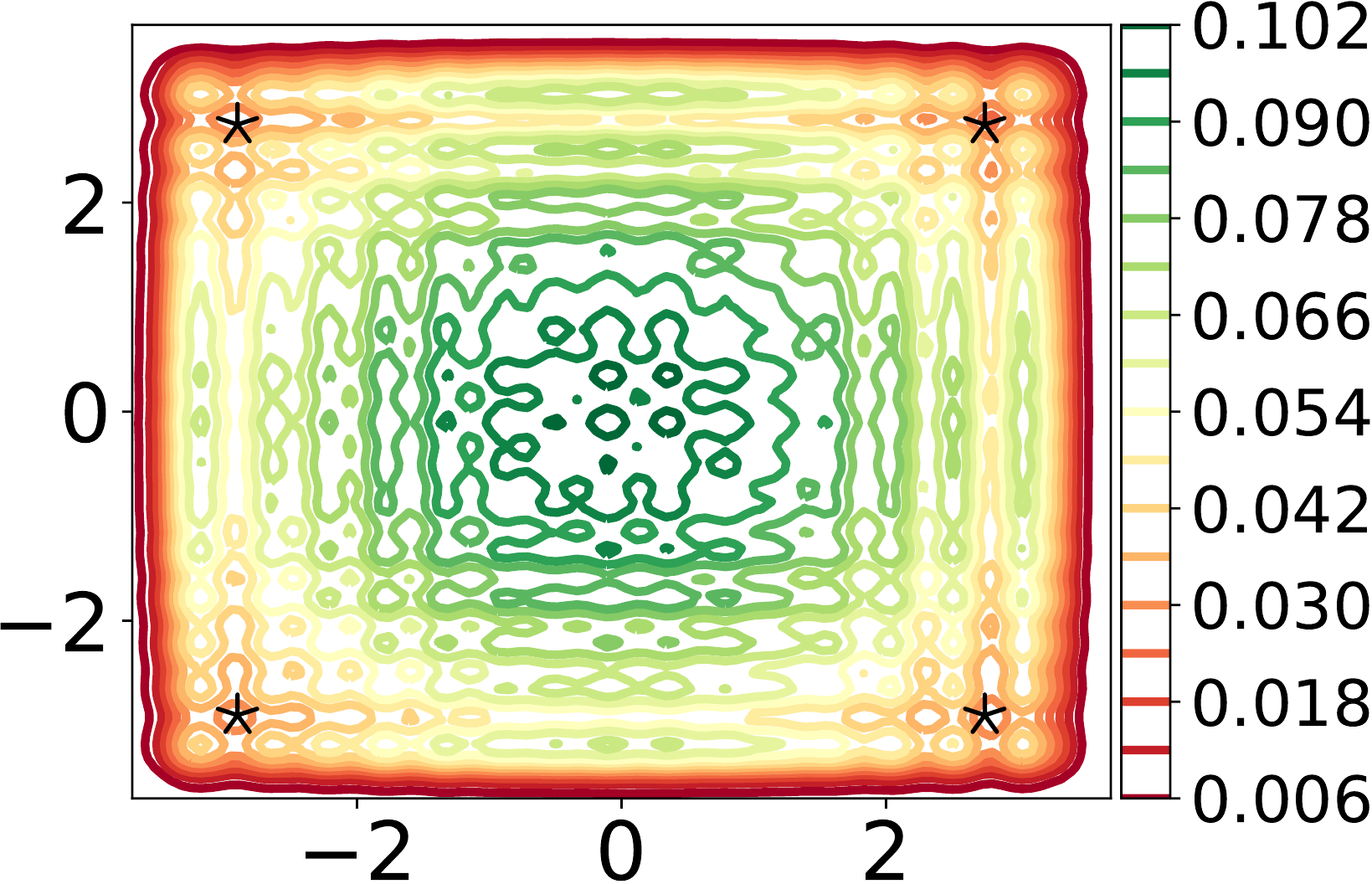}&
\includegraphics[width=0.4 \columnwidth]{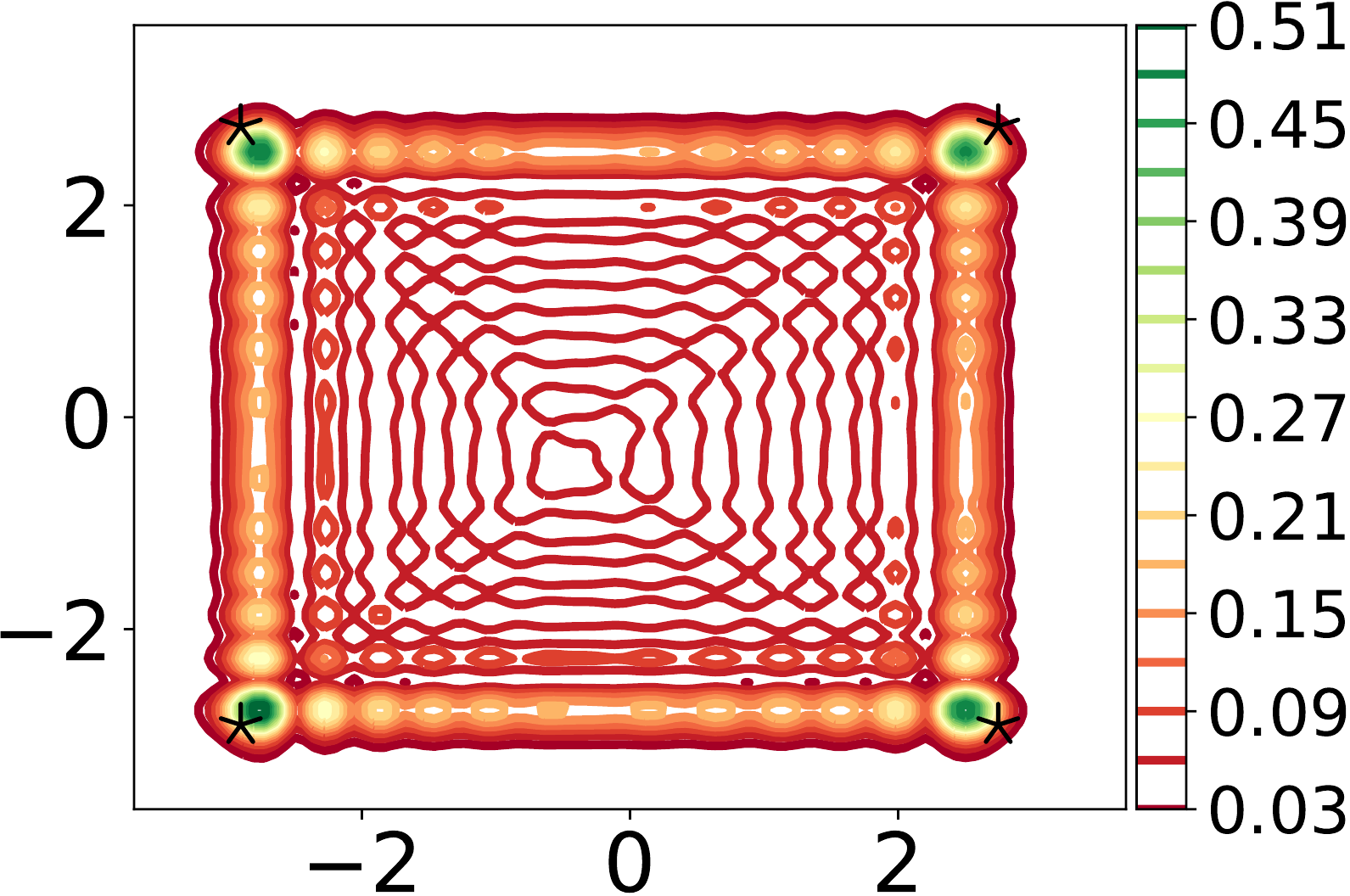}&
\includegraphics[width=0.4 \columnwidth]{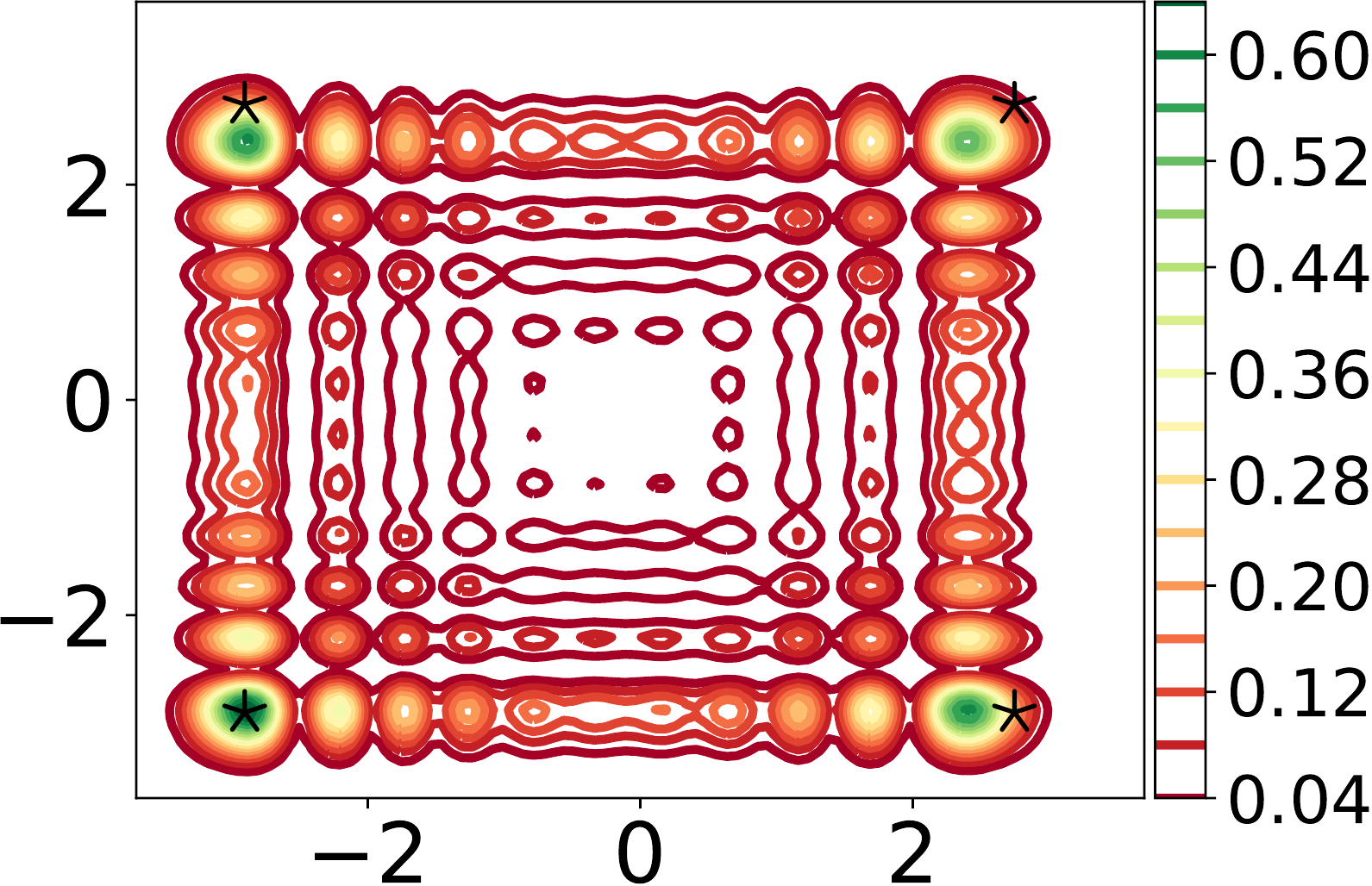}
\includegraphics[width=0.4 \columnwidth]{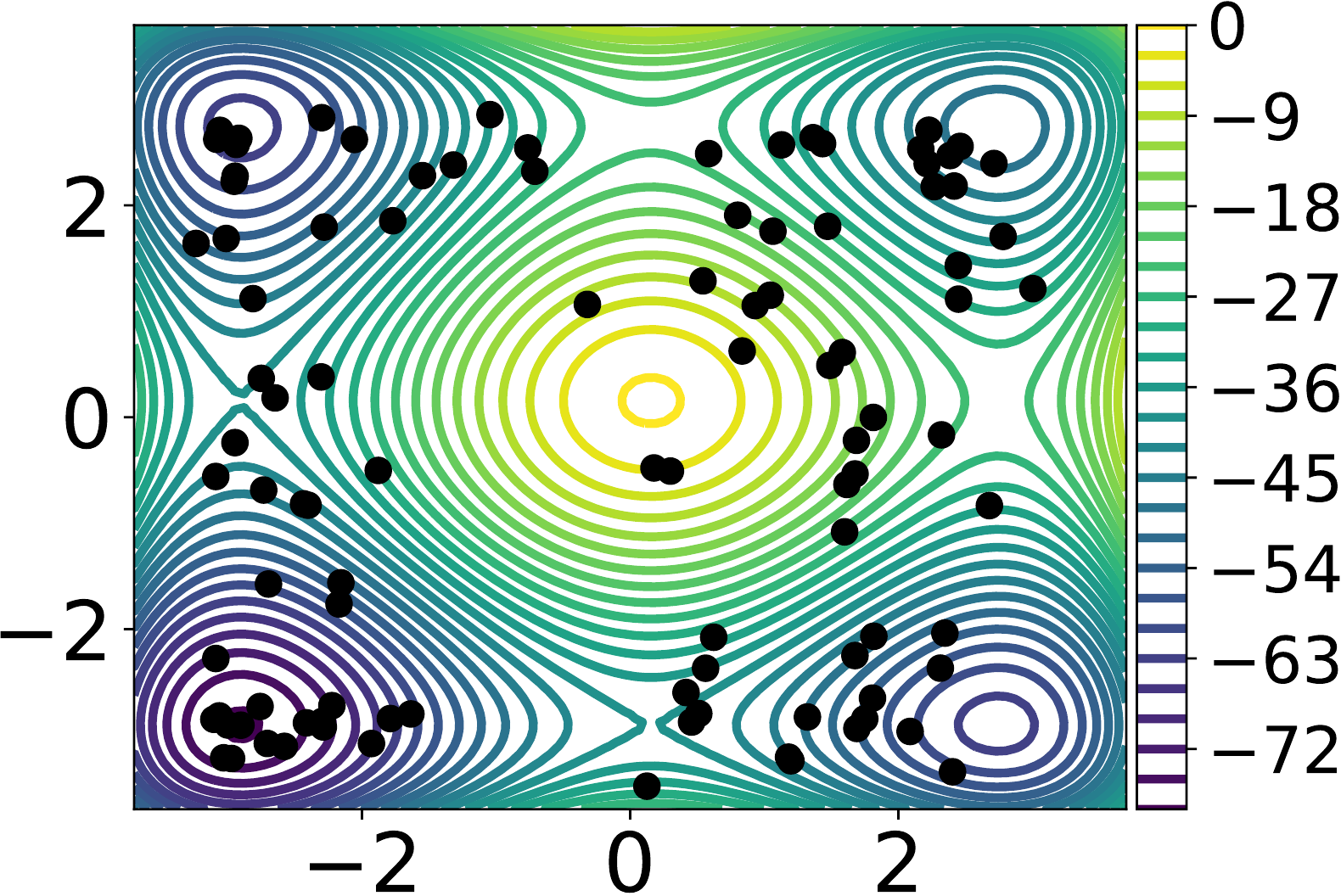}
\end{tabular}
\caption{Output probability distribution during subsequent steps of QAOA for optimizing the two-dimensional Styblinski-Tang (ST) function $f(x, y)=\frac 12(x^4-16x^2+5x+y^4-16y^2+5y)$. Note that the function is symmetric over $x,y$ so axes labels are not needed. The total number of steps in the algorithm is $P=3$. The local minima of the function are shown as black stars, with the global minimum located at $(x,y)=(-2.90353, -2.90353)$. The initial state is a momentum-squeezed two-mode vacuum, with squeezing parameter $r=1.0$ (9 dB). After a single step of the algorithm, interference fringes become visible and the distribution begins to spread towards regions of low function values. After two steps, the peaks of the distribution are concentrated around the minima, and this improves further after the final step, with the mode of the distribution located approximately at the global minimum.  On the right are the results of sampling from probability distribution of the output state of the algorithm. In the background, we show a contour plot of the two-dimensional ST function. The first 50 samples from the output state are shown in the foreground, with the best sample corresponding to $(x^*,y^*)=(-2.9027, -2.9066)$, which is close to the optimum $(-2.90353, -2.90353)$.}\label{Fig: ProbDbns}
\end{figure*}
\end{center}
\vspace{-2.9em}
\quad
Suppose that we have a target value $\bm{x}_f \in \mathbb{R}^N$ with corresponding oracle function $f(\bm{x}) = \delta^N(\bm{x}-\bm{x}_f)$. Following Ref.~\cite{pati2000quantum}, this function can be mapped to a cost Hamiltonian by defining the indicator displaced squeezed state $\ket{s_{\bm{x}_f}} := 
\tfrac{1}{(2 \pi \epsilon)^{n/4}} \int\mathrm{d}\bm{x}
\exp{[- \tfrac{(\bm{x} -\bm{x}_f)^2 }{4 \epsilon^2} ]}
\ket{ \bm{x} }$ and setting $\hat{H}_C=\ket{s_{\bm{x}_f}}\!\bra{s_{\bm{x}_f}}$. The exponential of this cost Hamiltonian can be done with quantum state exponentiation \cite{lloyd2014quantum,kimmel2017hamiltonian, rebentrost2018quantum}. As for the mixer, we choose a projector onto a state that is sharp in the momentum basis, namely $\hat{H}_M = \mathcal{F}^\dagger \ket{s_{\bm{x}_0}}\!\bra{s_{\bm{x}_0}} \mathcal{F}$, where $\mathcal{F}$ is the tensor product of Fourier transforms on all modes and $\bm{x}_0$ is some choice of initial momentum value. The resulting QAOA unitary $U(\bm{\nu},\bm{\gamma}) = \prod_{j=1}^P e^{-i\gamma_j \hat{H}_M} e^{-i\nu_j \hat{H}_C}$ reproduces Grover's algorithm and its speedup by setting $\gamma_j = \nu_j = \pi,\,\forall j$ and $P \sim \mathcal{O}(\sqrt{\pi^N})$. 

The inability of classical computers to simulate quantum computations over continuous variables has been most fruitfully studied in the context of CV-IQP circuits \cite{douce2017iqp,arrazola2017quantum}. These circuits have the following general structure: (i) inputs are multi-mode momentum-squeezed vacuum states, (ii) unitary operations $\hat{U}_f=e^{-if(\bm{\hat{x}})}$ are diagonal in the position basis, (iii) measurements are homodyne momentum measurements. By using Fourier transforms to map between position and momentum measurements, the output state of a CV-IQP circuit can be written as $\ket{\Psi} = \mathcal{F}e^{-if(\bm{\hat{x}})}\ket{\sigma_{\bm{p}}}= e^{i\bm{\hat{n}}\pi/4}e^{-if(\bm{\hat{x}})}\ket{\sigma_{\bm{p}}}$, where $\ket{\sigma_{\bm{p}}}$ is a momentum-squeezed vacuum state with squeezing level $\sigma$. This corresponds to a single step of a QAOA circuit with cost Hamiltonian $\hat{H}_C=f(\bm{\hat{x}})$, and a number mixer $\hat{H}_M=\bm{\hat{n}}$ with parameters $\gamma=-\pi/4$, $\nu=1$.  It has been shown in Refs.~\cite{douce2017continuous, arrazola2017quantum} that sampling from the output distribution of CV-IQP circuits in classical polynomial time is impossible unless the polynomial hierarchy collapses to third level. Thus, even for just one algorithmic step, the optimization procedure carried out by this variant of CV-QAOA cannot be duplicated efficiently with a classical device.

\textit{Numerical examples---}
Much can be understood about an optimization algorithm by testing it in practice. We consider the problem of minimizing the non-convex Styblinski-Tang function \cite{styblinski1990experiments}, a function often used for benchmarking optimization algorithms, which is defined as $f(\bm{x})=\frac 12\sum_{i=1}^N(x_i^4-16x_i^2+5x_i)$. The global minimum is located at $x_i=-2.90353$ for all $i=1,\ldots, N$. We focus on the two-dimensional case, allowing us to visualize the full output probability distribution and to classically simulate the algorithm with modest classical computation resources.     

The QAOA parameters are chosen such that $\nu_j=\gamma_j=T$ for all $j=1,\ldots,P$, so that only the parameter $T$ needs to be optimized. The guiding principle behind this choice is that, instead of reducing the learning rate to avoid overshooting during gradient descent, we iterate over different values of $T$ until the system reaches the minima after the final step. At this point the algorithm ends and no further descent takes place. The initial state is chosen as a momentum-squeezed two-mode vacuum, which corresponds to a large-variance Gaussian superposition over different initial positions. 

The output probability distribution -- as determined by the wavefunction of the output state -- is shown in Fig.~\ref{Fig: ProbDbns} for subsequent steps of the algorithm. In this example, we set $P=3$ and find good results by setting $T=0.1$. The simulation is carried out using the Strawberry Fields software for photonic quantum computing \cite{killoran2018strawberry}, with a custom quartic gate $e^{i \hat{x}^4}$ added to the library of built-in operations. Each mode is represented in the Fock basis, with a truncation of the Hilbert space to a cutoff of 90 photons. After the first step, it is already possible to identify a spreading-out of the probability density towards the local minima, just as expected from the gradient-descent interpretation of the algorithm. All points of the wavefunction cascade down their local gradient, and clear interference effects are influencing the overall distribution. After the second step, most of the distribution is concentrated around the four local minima. This effect becoming more pronounced during the final step, where the mode of the distribution is located approximately at the global minimum. 

Finally, we simulate sampling from the output state, as illustrated also in Fig.~\ref{Fig: ProbDbns}. From a total of 1000 samples, the best function value observed is $f(x^*, y^*)=-78.33216$, compared to the global optimum of $-78.3323$. Moreover, a total of 22 points were sampled for which the function value was smaller than $-78$, i.e., roughly 50 samples suffice on average to obtain good approximations to the global optimum. Indeed, samples are clustered around the four local minima, in accordance with the probability distribution of the output state shown in Fig.~\ref{Fig: ProbDbns}. 

\textit{Conclusion---}
We described a quantum algorithm that, based on the quantum dynamics of particles in energy potentials, can find approximate solutions to continuous optimization problems. The algorithm can be interpreted as performing gradient descent in superposition, a feature that is not only intriguing but also useful in searching for optimal parameters. The algorithm is also versatile: as we argued, it can in principle handle both constrained and discrete optimization problems. Moreover, in general, simulating its output distribution cannot likely be done efficiently using classical computers. 

The ability to encode Grover's search algorithm is an indicator of potential speedups achievable with this method. Nevertheless, as with other quantum approximate optimization algorithms, it is crucial to determine whether it can offer an advantage compared to existing classical techniques for specific problems of practical interest. Finally, it is important to note that implementing the algorithm may require technologies beyond those currently available, either with the advent of continuous-variable quantum computers or by employing large numbers of qubits capable of approximating continuous degrees of freedom.

\textit{Acknowledgements---}
The authors would like to thank Jason Pye and Christian Weedbrook for useful discussions, as well as Micha\l\ St{\k e}ch{\l}y for originally inquiring about a continuous version of QAOA. GV acknowledges funding from NSERC.
This research was supported in part by Perimeter Institute for Theoretical Physics. Research at Perimeter Institute is supported by the Government of Canada through the Department of Innovation, Science and Economic Development and by the Province of Ontario through the Ministry of Research and Innovation.

\bibliographystyle{apsrev}

\bibliography{Bibliography}

\end{document}